# Models for Truthful Online Double Auctions

**Jonathan Bredin**
Department of Mathematics
Colorado College
Colorado Springs, CO 80903
bredin@acm.org

**David C. Parkes**
Division of Engineering and Applied Sciences
Harvard University
Cambridge, MA 02138
parkes@eecs.harvard.edu

## Abstract

Online double auctions (DAs) model a dynamic two-sided matching problem with private information and self-interest, and are relevant for dynamic resource and task allocation problems. We present a general method to design truthful DAs, such that no agent can benefit from misreporting its arrival time, duration, or value. The family of DAs is parameterized by a pricing rule, and includes a generalization of McAfee's truthful DA to this dynamic setting. We present an empirical study, in which we study the allocative-surplus and agent surplus for a number of different DAs. Our results illustrate that dynamic pricing rules are important to provide good market efficiency for markets with high volatility or low volume.

## 1 INTRODUCTION

We consider the problem of matching self-interested parties in dynamic multi-agent systems. Cast as a task-allocation problem, each agent arrives in some period and is present in the system for some finite period of time. An agent is either a seller, able to perform a single job while present, or a buyer, with a job to complete. All jobs are identical, and any present agent can perform a job. All jobs are completed in a unit period of time. The arrival, departure, and cost (if a seller) or value (if a buyer) are all private information to agents. Moreover, each agent is assumed to be self-interested and willing to misreport its private information (including arrival and departure times) if this can improve the outcome in its favor.

We model this problem as an online double auction for identical items. The matchmaker becomes the auctioneer. For dynamic task allocation, an item represents a task to be performed. Each agent is only willing to buy (or sell) a single unit of the item. Uncertainty about the future coupled with the two-sided nature of the market leads to an interesting mechanism design problem.

For example, consider the scenario where the auctioneer must decide how (and whether) to pair an immediately departing seller with reported cost of $6 with buyers, one of which has a reported value of $8 and one a reported value of $9. Should the auctioneer pair the higher bidder with the seller? What happens if a seller willing to sell for $4 arrives after the auctioneer acts upon the matching decision? How should the matching algorithm be designed so that no agent can benefit from misstating its private constraints on arrival and departure, and its private cost or value information?

We introduce a general framework for the design of *truthful* double auctions (DAs) in this dynamic model. The DAs are *truthful* in the sense that the optimal strategy for an agent, whatever the future auction dynamics and whatever the bids from other agents, is to report its true value (or cost for sellers) and true departure period immediately upon arrival into the auction. We obtain auctions that are truthful in a *dominant-strategy equilibrium* (DSE). This is useful because it simplifies the decision problem facing bidders. An agent can determine its optimal bidding strategy without a model of either the auction dynamics or the other agents.

The main technical challenge is to provide truthfulness while also ensuring (weak) budget-balance. We require that DAs do not run at a deficit, i.e. at no point in time should the auctioneer have paid more than it has received. In developing a family of truthful auctions we apply the price-based characterization from Hajiaghayi *et al.* [9] to develop an implementable price rule that satisfies the required properties of *agent-independence* and *monotonicity*. This ensures truthfulness in a DSE. To prevent the auction running at a deficit, the individualized price faced by each agent to trade must increase across time. We make no such assumption, however, on the global price schedule, for instance as in Lavi and Nisan [13], because we study an infinite time horizon setting that continuously runs a DA.

Truthfulness in DAs also requires new elements, for instance an agent's bid value can only influence the price faced by other agents once it is determined whether the agent will trade or not. This is important because the *availability* of trades depends on the price faced by other agents. For example, a buyer that is required to pay $4 in the DA to trade might like to decrease the price that a potentially matching seller will receive from $6 to $3 to allow for trade. In this aspect, our approach generalizes the truthful DA of McAfee [14] to the online setting.

Our main contribution is the characterization of a family of parameterized truthful DAs. Different price schedules can be substituted and used within matching. From within this class we describe a fixed price scheme, exponentially-weighted moving average and windowed-average schemes, and a scheme based on McAfee's rules for a static DA [14]. An empirical analysis illustrates the market efficiency of each auction for different market conditions. For high-volume and low volatility markets, setting a single well-chosen price is reasonably effective in supporting efficient trades. However, when volume is lower or when demand is volatile, then DAs with dynamic pricing rules, perhaps moving-average or McAfee-based, are most efficient.

## 1.1 RELATED WORK

Static two-sided market problems have been widely studied [15, 4, 17, 21, 7]. In a classic result, Myerson and Satterthwaite [15] proved that it was impossible to achieve efficiency with voluntary participation and without running a deficit, even relaxing DSE to a Bayesian-Nash equilibrium.[1]

Our problem is also similar to a traditional *continuous double auction* (CDA), where buyers and sellers may at any time submit offers to a market that pairs an offer as soon as a matching offer is submitted. Early work considered market efficiency of CDAs with human experiments in labs [18], while recent work investigates the use of software agents to execute trades [19]. While these markets have no dominant strategy equilibria, populations of software trading agents can learn to extract virtually all available surplus, and even simple automated trading strategies outperform human traders [6]. However, these studies of CDAs assume that all traders share a known deadline by which trades must be executed. This is quite different from our setting, in which we have dynamic arrival and departure.

Some truthful DAs are known for static problems [14, 11, 2, 1]. For instance, McAfee [14] introduced a DA that sometimes forfeits trade in return for achieving truthfulness. Truthful one-sided online auctions, in which agents arrive and depart across time, have received some recent attention [13, 10, 9, 16, 12]. We adopt and extend the monotonicity-based truthful characterization in the work of Hajiaghayi et al. [9] in developing our framework for truthful DAs. Our model of DAs must also address some of the same constraints on timing that occur in Porter [16] and Hajiaghayi *et al.* [9] and Lavi and Nisan [12]. In these previous works, the items were *reusable* or *expiring* and could only be allocated in particular periods. In our work we provide limited allowance to the matchmaker, allowing it to hold onto a seller's item until a matched buyer is ready to depart (perhaps after the seller has departed).

Blum *et al.* [3] present online matching algorithms whose competitive ratios depend on the spread in offer valuations. This earlier work does not consider incentive-compatibility, and is presented purely from an algorithmic perspective.

## 2 PRELIMINARIES

In our model of online DAs we consider discrete time periods $T = \{0, 1, 2, \ldots\}$, with buyers and sellers that arrive and depart over time are each interested in trading a single unit of an identical good.

An agent $i$ has *type*, $v_i = (a_i, d_i, r_i, w_i) \in V$, where $r_i \in \{b, s\}$ indicates whether an agent is a buyer ($b$) or a seller ($s$), $a_i \in T$ is the arrival time, and $d_i \in T$, with $d_i \geq a_i$, denotes the departure time. For a buyer, $w_i \in \mathbb{R}_{\geq 0}$ defines its value for receiving one unit of the item while present. For a seller, $w_i \in \mathbb{R}_{\leq 0}$ defines its value for selling one unit of the item while present.[2][3] We interpret an agent's arrival time as either the time at which it first becomes aware of the auction, or the time at which it first becomes aware of its desire to trade. For a buyer, we interpret the departure period as the final period in which it values the item. For a seller, the departure period is the final period in which it is willing to receive payment.

---

[1] In a Bayesian-Nash equilibrium each strategy must maximize the expected utility of an agent given its beliefs about the types of the other agents.

[2] This simple "step-function" model of temporal preferences, in which an agent is indifferent for some period of time and then has no value, is adopted throughout the current literature on truthful online auctions.

[3] In general, we use the term *bid* to refer to offers from both buyers and sellers. Sometimes we need to refer to offers from the buy-side or sell-side in particular, in which case we will reserve *bid* for buyers and *ask* for offers from sellers.

From the example in the introduction, the two buyers could have types $v_1 = (1, 4, b, 8)$ and $v_2 = (2, 5, b, 9)$, and the seller could have type $v_3 = (2, 3, b, -6)$. At time 3, the auctioneer may consider pairing $v_3$ with one of the two bids, since either of the two buyers will meet $v_3$'s ask price. The value for trades in all other periods is zero for buyers and $-\infty$ for sellers, indicating that no agent will trade outside of its patience interval.

Agents are assumed to be self-interested, and their types are private information. The truthful DAs that we model are *direct-revelation*, meaning that an agent bids by making a claim about its type, i.e. reporting type $\hat{v}_i = (\hat{a}_i, \hat{d}_i, \hat{r}_i, \hat{w}_i) \neq v_i$ to the auction.[4] An agent's self-interest is exhibited in its willingness to misrepresent its type when this will improve the outcome of the DA in its favor (e.g. matching at a beneficial price when not matching with truthful reports, or matching at a better price.)

We make the following assumptions:

**(A1)** Agents cannot understate their arrival period, i.e. $\hat{a}_i \geq a_i$.

**(A2)** All agents have a bounded patience, with $d_i \leq a_i + K$, for all $v_i \in V$.

**(A3)** The auctioneer can "hold" an item that is matched between a seller-buyer pair, releasing it when the buyer is ready to depart.

**(A4)** No false-name bids, and no collusion.

Assumption (A1) is standard and is consistent with our interpretation of the arrival time. Other misreports are allowed in the model. Assumption (A2) is novel, but required for our family of mechanisms because it is enables the definition of an individualized, yet increasing price, to each agent. Assumption (A3) is required to achieve truthfulness with agents that are able to over-state their departure, because it allows the appropriate timing of trades.[5] We need assumption (A4), no false-name bids [20] (submitting multiple bids as a single bidder) and no collusion, because in both cases an agent could influence the price it faces or the price to agents on the other side of the market.

### 2.1 DESIDERATA

A DA is defined as a pair, $M = (g, \tilde{p})$, with *allocation rule*, $g$, and *payment rule*, $\tilde{p}$. The allocation rule, $g_i(\hat{v}_{\leq t}, t) = 1$, indicates that agent $i$ buys an item in period $t$, given bids $\hat{v}_{\leq t} = \{\hat{v}_j = (\hat{a}_j, \hat{d}_j, \hat{w}_j) : \hat{a}_j \leq t\}$, and $g_i(\hat{v}_{\leq t}, t) = -1$ indicates that agent $i$ sells an item in period $t$. The payment rule $\tilde{p}_i(\hat{v}_{\leq t}, t) \in \mathbb{R}$ indicates a payment made by agent $i$ to the auctioneer in period $t$ (this is a payment *from* the auctioneer to the agent when $\tilde{p}_i(\hat{v}_{\leq t}, t) < 0$). We need $g_i(v_{\leq t}, t) \geq 0$ for all buyers, with $g_i(v_{\leq t}, t) = 1$ in at most one period $t \in [a_i, d_i]$ and zero otherwise (and similarly for sellers, with $g_i(v_{\leq t}, t) \leq 0$ and $g_i(v_{\leq t}, t) = -1$ in at most one period $t \in [a_i, d_i]$.) Payments can only be made (or collected) when an agent is present.

---
[4]This is without loss of generality, since the revelation principle continues to hold for online mechanisms [8].

[5]In a task-allocation setting, this assumption is interpreted as allowing the matchmaker to hold onto the result of completing the task (e.g. some computational result) until the matched buyer is ready to depart.

We assume that agents have *quasi-linear* utility functions, so that the utility for a trade $x$ at some price $p$ is the value from the trade minus the payment. Agents are modeled as utility-maximizing. Given assumption (A1), we adopt $L(v_i) \subseteq V$ to denote the set of possible reports, given true type $v_i$. Set $L(v_i)$ contains all types $\hat{v}_i \in V$ for which $\hat{a}_i \geq a_i$. We can now define truthfulness. Truthfulness requires that no misreport can *ever* improve an agent's utility, whatever the reports of other agents and whatever the future dynamics:

**Definition 1 (Truthfulness).** *A mechanism $M = (g, \tilde{p})$, where $g$ is an allocation rule and $\tilde{p}$ is a payment rule, is truthful if for any agent $i$ and any $v_i$, we have $v_i(g_i(v)) - \tilde{p}_i(v) \geq v_i(g_i(\hat{v}_i, v_{-i})) - \tilde{p}_i(\hat{v}_i, v_{-i})$, for all $\hat{v}_i \in L(v_i)$ and all $v_{-i}$.*

Here, allocation $g_i(v) = \sum_{t' \in [a_i, d_i]} g_i(v_{\leq t'}, t')$, and indicates whether agent $i$ gets the item in any period, $g_i(\hat{v}_i, v_{-i})$ is shorthand for "the allocation for agent $i$ given that the bids from other agents are $v_{-i}$ and agent $i$ bids $\hat{v}_i$," and value $v_i(x) = w_i$ if $x = 1$ and 0 otherwise. Payment $\tilde{p}_i(v) = \sum_{t \in [a_i, d_i]} \tilde{p}_i(v_{\leq t}, t)$, and is the total payment across all periods in its arrival departure interval, and $\tilde{p}_i(\hat{v}_i, v_{-i})$ is shorthand for "the payment by agent $i$ given that the bids from other agents are $v_{-i}$ and agent $i$ bids $\hat{v}_i$."

In addition to truthfulness, we adopt the following as required properties of DAs:

**Definition 2 (No-Deficit).** *A mechanism $M = (g, \tilde{p})$ runs at no-deficit if:*

$$\sum_{i: d_i < t} \tilde{p}_i(v) + \sum_{i: a_i \leq t \leq d_i} \sum_{t' \in [a_i, t]} \tilde{p}_i(v, t') \geq 0, \quad \forall t, \forall v. \quad (1)$$

No-deficit requires that the auctioneer has money on hand in each time period.

**Definition 3 (Feasible Trade).** *A mechanism $M = (g, \tilde{p})$ is feasible if:*

$$\sum_{i: d_i < t} g_i(v) + \sum_{i: a_i \leq t \leq d_i} \sum_{t' \in [a_i, t]} g_i(v, t') \geq 0, \quad \forall t, \forall v. \quad (2)$$

Feasibility requires that the auctioneer has physical goods on hand in each time period. This still provides flexibility to the auctioneer to hold onto items until a matched buyer is ready to depart.

## 3 PRICE-RANKED DOUBLE AUCTIONS

We define a family of truthful, no-deficit and feasible online DAs. The DAs are parameterized by a price schedule, $f_i$, that when coupled with a simple matching rule defines an allocation rule $g$ and payment rule $\tilde{p}$. In Section 4 we prove that any DA for which the price schedule $f_i$ is *valid* will be truthful.

At the start of period $t$, bids in the auction that are within their patience interval (i.e. $t \in [a_i, d_i]$) can be in one of three states: *matched, priced-out*, or *active*. A matched bid will definitely trade and its price has been determined. A bid is priced-out when the auction has determined that it will definitely not trade. An active bid is neither priced-out nor matched, and whether or not it will trade remains uncertain.

For all bids and all periods $t$ while a bid is active or not yet arrived, we define a valid price schedule $f_i(t, r_i, v_{-i}) \in \mathbb{R}$ to determine the price to buy (with $r_i = b$) or sell (with $r_i = s$) an item in period $t$ given bids $v_{-i}$ from other agents.[6]

**Definition 4 (Valid price schedule).** *A price-schedule, $f$, is* valid *if it satisfies each of the following three properties in any period $t$ before a bid is priced-out or matched:*

**(B1)** *Price $f_i(t, r_i, v_{-i})$ does not depend on the arrival time, departure time, or on the reported value, of bidder $i$.*

**(B2)** *Price $f_i(t, r_i, v_{-i})$ does not depend on bids that arrive after period $t$.*

**(B3)** *While $f_i(t, r_i, v_{-i}) \leq w_i$ the price does not depend on the value of any bids $j \neq i$ that are active at the start of period $t$ and have $f_j(t, r_j, v_{-i}) \leq w_j$. Otherwise, if $f_i(t, r_i, v_{-i}) > w_i$ then we also have $f_i(t, r_i, v_{-i}) > w_i$ for all values of bids $j \neq i$ that are active at the start of period $t$ while $f_j(t, r_j, v_{-i}) \leq w_j$.*

Property (B1) states *agent independence* and requires that the price is independent of all elements of an agent's reported type except whether or not it claims to be a buyer or a seller. Property (B2) asserts on-line computability.

Property (B3) limits the way in which a price to agent $i$ can depend on the bids from *other* agents. This property is novel to online DA, and required

---

[6]This price is best thought of as a lower bound on the final payment made, and is only exact if all agents are impatient (i.e. for $K = 1$).

---

to achieve truthfulness in combination with budget-balance. Note that the price can depend in arbitrary ways on information in bids that have departed, matched in a previous period, or priced-out. Informally, property (B3) states "*if you might still trade then your bid should not affect the price of other bids that might still trade or whether or not other bids are priced-out.*" First, if bid $i$ is priced below its value, with $f_i(t, r_i, v_{-i}) \leq w_i$, then its price must be independent of the reported value of all bids active at the start of the period and with a price in this period that allows them to remain active. Second, if bid $i$ is priced above its value then it must be priced-out whatever the bid values from other bids.

In the next section we define simple price schedules that satisfy (B1)–(B3). For now, we define the matching function that completes the definition of our class of DAs. The **price-ranked DAs**, $M_{\text{pr}}(f)$, define a provisional price, $\tilde{ps}_i(t)$, to agent $i$ in period $t$, as:

$$\tilde{ps}_i(t) = \max_{\tau \in \{d_i - k, \ldots, t\}} \{f_i(\tau, r_i, v_{-i})\}, \quad (3)$$

for all periods $t$ until the bid is matched or departs. Here, the dependence on the reported type of the agent and on the reports of other agents is left implicit.

**Definition 5 (Price-Ranked Matching Rule).** *At the start of each period $t$, the matching rule considers those bids that are still* active. *For these bids,*

1. *Compute the new price $f_i(t, r_i, v_{-i})$ to each active bid. Mark all active bids for which $f_i(t, r_i, v_{-i}) > w_i$ as* priced-out. *Update $\tilde{ps}_i(t) = \max\{\tilde{ps}_i(t-1), f_i(t, r_i, v_{-i})\}$ for the remaining active bids.*

2. *Sort the active bids into a bid book, in order of decreasing price $f_i(t, r_i, v_{-i})$. Sort the active asks into an ask book, in order of decreasing price $f_j(t, r_j, v_{-j})$. Break ties at random.*

3. *Match bids and asks while $f_i(t, r_i, v_{-i}) + f_j(t, r_j, v_{-j}) \geq 0$, matching from the top of the bid book and the ask book. Set the payment to a matched bidder as $\tilde{p}_i(v) = \tilde{ps}_i(t)$.*

*Once matched, a trade between a pair of agents is timed as follows: if the seller departs first, the buyer makes payment at this time and the seller provides the item to the auctioneer, which releases the item at the departure of the buyer; if the buyer departs first, the seller provides the item to the buyer at this time and the buyer pays the auctioneer which releases the payment at the departure of the seller*

As an example, assume that we have a price schedule $f_i(t, r, v_i)$ with which to match offers and that the patience bound $K$ is 5. A bid $v_i = (10, 13, b, 6)$ arrives. The auctioneer must compute a provisional price using Equation 3 over periods 8, 9, and 10. Say $f_i(8, b, v_{-i}) = 7$, $f_i(9, b, v_{-i}) = 5$, and $f_i(10, b, v_{-i}) = 4$. Then $v_i$ is marked as priced-out and will never be considered for matching. If, however the bidder is more patient and $v_i = (10, 14, b, 6)$, then the auctioneer places $v_i$ into the bid book sorted by $f_j(10, b, v_{-j})$. The auctioneer matches offers from the front of the bid and ask books, matching bid $i$ with ask $j$ only if doing so would yield budget balance at the *current* price– if $f_i(b, 10, v_{-i}) + f_j(s, 10, v_{-j}) \geq 0$. In that event, bidder $i$ pays 5, the maximum price faced over its theoretical lifetime. If $v_i$ does not match in period 10, then it is kept until it expires, sorted in the bid book by the most recent $f_i(t, b, v_{-i})$. The auctioneer removes $v_i$ either when $v_i$ trades, expires, or is priced out ($w_i = 6 < f_i(t, b, v_{-i})$).

## 3.1 VALID PRICE SCHEDULES

The family of truthful DAs is defined in terms of valid price schedules. The main constraint for a price schedule to be valid is that the price in any period must be independent of the active bids in that period. In this section, we present three schedules: fixed price, moving average (exponential and windowed), and McAfee-based. The price-ranked DAs defined for each schedule are evaluated in Section 5.

### 3.1.1 Fixed Pricing

A basic price schedule sets a fixed price, with

$$f_i^{\text{fixed}}(t, r_i, v_{-i}) = \begin{cases} p_b^* & \text{if, } r_i = b \\ p_s^* & \text{if, } r_i = s \end{cases}$$

and $p_b^* \geq 0$ and $p_s^* \leq 0$ set to maximize efficiency given prior beliefs of the auctioneer (or via experimentation), and generally with $p_b^* = -p_s^*$. The fixed price schedule trivially satisfies (B1)–(B3) by construction. A fixed-price schedule is shown to perform reasonably when the market is predictable and when volume is high.

### 3.1.2 Moving Average

The fixed-price schedule is not appropriate when market conditions fluctuate. Instead, we can adapt an exponentially-weighted moving average on some statistic of the offers that have expired, priced-out or matched in the current period. Alternatively, we can compute a similar statistic on a rolling window of these bids.

An *exponentially-weighted moving average price* is defined for buyers and sellers, respectively, as

$$f_i^{\text{ewma}}(t, b, v_{-i}) = \\ \lambda \, \xi(v_{t-1}) + (1-\lambda) f_i^{\text{ewma}}(t-1, b, v_{-i}),$$
$$f_i^{\text{ewma}}(t, s, v_{-i}) = \\ -\lambda \, \xi(v_{t-1}) + (1-\lambda) f_i^{\text{ewma}}(t-1, s, v_{-i}),$$

where $\lambda \in [0, 1]$ is a smoothing constant. Here, $\xi(v_t)$, is a statistic defined on the values of offers (bids and asks) that expire, are priced-out, or trade in period $t$, denoted $v_t$. We find that the *mean statistic*, $\xi^{\text{mean}}(v_t)$, of the absolute values of offers performs well.

An alternate approach computes the price schedule from a statistic over a fixed-size window of the most recent expired, priced-out, and traded offers. The price is simply $f_i^{\text{window}}(t, b, v_{-i}) = \xi(v_{t-1,\phi})$ for buyers and $f_i^{\text{window}}(t, s, v_{-i}) = -\xi(v_{t-1,\phi})$ for sellers, where we update the meaning of $v_{t,\phi}$ to represent the set of the most recent $\phi$ expiring, priced-out, and traded offers up to time $t$. We find that two statistics work well: the **median** of absolute values of offers, and a simple calculation that sets a **clearing price** to maximize the number of matching offers.

For cases in which $\xi(v_{t-1})$ is not well-defined, usually due to low volume leading to missing data because no offers are in the sample set, we set $f_i(t, r_i, v_{-i}) = f_i(t-1, r_i, v_{-i})$ in both the exponentially-weighted and windowed schemes. Both moving averages implement valid price schedules, satisfying (B1) and (B3) since no bids that remain active in the current period are used for pricing.

The main concern in setting prices is that they are too volatile. Fluctuating prices make the constraint $w_i \geq \tilde{p}s(t)$ more difficult to satisfy, since provisional prices are driven upwards through Equation (3). Price fluctuation also creates a bid-ask price spread since prices to both buyers and sellers are driven higher, so that buyers pay more while sellers pay less (their payments are less negative). Smoothing and windowing are useful for both of these reasons: providing long-term learning while dampening short-term variations. The choice of smoothing factor, $\lambda$, or window size can be determined experimentally.

### 3.1.3 McAfee Rule

We can also adopt the pricing method in McAfee's DA [14] to define a valid price schedule for our online auction setting. The main difference with the previous methods is that it uses current bids to price trades. This can help to make it more responsive.

McAfee's DA is useful for the static problem of pairing bids and asks present in a single time period. The best way to understand McAfee's clearing and pricing rule is through a simple example. Consider 4 bids, with values $(10, 6, 4, 2)$ and 4 asks, with values $(-2, -4, y, -12)$, with $y \in \{-6, -10\}$. Let $m$ denote the index of the last bid to clear, i.e. $b(m) + s(m) \geq 0$ and $b(m+1) + s(m+1) < 0$, with $m = 2$ in our example. McAfee's rule is defined on two cases:

- (Case I) If there are at least $m+1$ bids and asks and price $p_{m+1} = (b(m+1) - s(m+1))/2 \leq b(m)$ and $-p_{m+1} \leq s(m)$ then the first $m$ bids and asks trade at price $p_{m+1}$ for bids and $-p_{m+1}$ for asks.
- (Case II) If there are fewer than $m+1$ bids or asks, or the price test in (Case I) fails, then $m-1$ bids and asks trade at price $b(m)$ for bids and price $s(m)$ for asks.

In our example, when $y = -6$ we are in Case I and $p_{m+1} = 5$, but for $y = -10$ we are in Case II and $p_{m+1} = 7$ and so only the first bid and ask trade, at prices 6 and $-4$ respectively. McAfee's auction is truthful and does not run at a budget deficit, and implements the efficient trade in Case I but forfeits one trade in Case II.

We can now interpret McAfee's auction in terms of an agent-independent price function. For this, for bid $i$, let $mb_i$ and $ms_i$ denote the last bid and ask to clear (in the sense of index $m$ above) when bid $i$ is removed along with the maximal ask on the other side of the market. For ask $j$, let $mb_j$ and $ms_j$ denote the last bid and ask to clear when ask $j$ is removed along with the maximal bid on the other side of the market. Consider the example, with $y = -10$. For bid 2, we have $mb_2 = 1$ with $b(mb_2) = 10$ and $ms_2 = 2$ with $b(ms_2) = -4$. For ask 3, we have $mb_3 = 3$ with $b(mb_3) = 4$ and $ms_3 = 2$ with $s(ms_3) = -4$.

Now, McAfee's DA can be understood in terms of the following agent-independent prices, which we will adopt for a price schedule.[7] Let $p(i) = (b(mb_i + 1) - s(ms_i + 1))/2$. For a bid $i$, we have:

$$f_i^{\text{mcafee}}(t, b, v_{-i}) = \begin{cases} p(i) & \text{if } p(i) \leq b(mb_i) \\ & \text{and } p(i) \geq -s(ms_i) \\ b(mb_i) & \text{otherwise.} \end{cases}$$

For an ask $j$, we have:

$$f_j^{\text{mcafee}}(t, s, v_{-j}) = \begin{cases} -p(j) & \text{if } -p(j) \leq s(ms_j) \\ & \text{and } -p(j) \geq -b(mb_j) \\ s(ms_j) & \text{otherwise.} \end{cases}$$

Consider the example, with $y = -6$. For bid 2, we have $p(2) = (4 - (-6))/2 = 5$ and $p(2) = 5 \leq b(mb_2) = b(1) = 10$ and $p(2) = 5 \geq -s(ms_2) = -s(2) = 4$, and so $f_2^{\text{mcafee}}(t, b, v_{-2}) = p(2) = 5$. For bid 3, we have $p(3) = (2 - (-12))/2 = 7$ and $p(3) = 7 > b(mb_3) = b(2) = 6$ and so adopt $f_3^{\text{mcafee}}(t, b, v_{-3}) = b(mb_3) =$

---

[7] When every bid (or every ask) clears then we to make McAfee pricing well-defined we must also adopt 0 for missing bid prices $b(mb_i + 1)$ or $b(mb_j + 1)$, and adopt $-\infty$ for missing ask prices. Similar special cases can also be handled when one or more of $mb_i, ms_i, mb_j$ or $ms_j$ are undefined.

$b(2) = 6$.[8] The proof of the following is omitted in the interest of space.

**Proposition 1.** *The McAfee-based price schedule is valid, and $f_j^{\text{mcafee}}(t, s, v_{-j})$ equals the price in McAfee's DA in the current period for any agent $j$ that would trade in McAfee's DA in the current period.*

We can make the following immediate observation (the proof is trivial and omitted in the interest of space):

**Proposition 2.** *The price-ranked DA with a McAfee-based price schedule reduces to a sequence of standard McAfee auctions when every bidder is impatient, i.e. with $K = 1$.*

In passing, we observe that a potential problem with McAfee in our online setting is that the price swings may be large, driving a a large bid-ask price spread through the max operator in Equation (3). However, applying smoothing (as in the exponentially-weighted moving average approach) in this case loses property (B3) because decisions about which agents are priced-out can be reversed via smoothing.

## 4 THEORETICAL ANALYSIS

In this section, we prove our main result which is that each auction in the family of price-ranked DAs with valid price schedules. First, we demonstrate the no-deficit and feasibility properties.

**Theorem 1.** *The price-ranked DA, $M_{\text{pr}}(f)$, is no-deficit and feasible.*

*Proof.* To establish no-deficit, just observe that the price-ranked DA only selects a bid-ask pair to match when the sum of the current prices $f_i(t, r_i, v_{-i}) + f_j(t, r_j, v_{-j}) \geq 0$. By Equation (3), we know that $\tilde{ps}_i(t) \geq f_i(t, r_i, v_{-i})$, for all agents and so checking for no-deficit in terms of the simple price schedule $f$ also implies no-deficit in terms of the provisional prices that become each agent's payment. Feasibility is immediate, because we only match bids and asks in pairs. □

To establish truthfulness for the price-ranked DAs we appeal to the price-based characterization in Haji-aghayi et al. [9]. This theory was constructed for one-sided online auctions, but continues to hold for DAs. Some care is required to handle the budget-balance requirement in online DAs in a way that is consistent with the properties required for truthfulness.

---

[8] The interested reader can check for this example that these prices define the payments in the McAfee rule for bids and asks that trade, and are priced above the value of bids or asks that do not trade. In this example, the buy-side prices are $(5, 5, 6, 6)$ for $y = -6$ and $(6, 7, 8, 8)$ for $y = -10$ and the sell-side prices are $(-5, -5, -4, -4)$ for $y = -6$ and $(-4, -2, -4, -4)$ for $y = -10$.

**Definition 6 (Price-based Auctions).** *An online auction is price-based if there exists a value-independent price schedule $p_i(a_i, d_i, r_i, v_{-i}) \in \mathbb{R}$, such that, for all valuations $v$:*

- *(admissible) For all $i$, if $\hat{r}_i = b$ then $g_i(v) = 1$ if and only if $p_i(a_i, d_i, b, v_{-i}) \leq w_i$, and if $\hat{r}_i = s$ then $g_i(v) = -1$ if and only if $p_i(a_i, d_i, s, v_{-i}) \leq w_i$.*
- *(payments) For all $i$, if $\hat{r}_i = b$ and $g_i(v) = 1$ then $\tilde{p}_i(v) = p_i(a_i, d_i, b, v_{-i})$, and if $\hat{r}_i = s$ and $g_i(v) = -1$ then $\tilde{p}_i(v) = p_i(a_i, d_i, s, v_{-i})$.*

Value-independent prices $p_i$ can depend on arrival, departure and buy/sell type, but not on the value of a bid. Property *(admissible)* states that an agent must trade whenever its value is larger than the price. Property *(payments)* states that the payment must be exactly the price faced by the agent.

The following property of prices is important in establishing truthfulness:

**Definition 7 (Monotonic).** *Prices are monotonic if $p_i(a_i, d_i, r_i, v_{-i}) \leq p_i(a'_i, d'_i, r_i, v_{-i})$ for all $a'_i \geq a_i$ and all $d'_i \leq d_i$, and all $v_{-i}$ and all $r_i$.*

In words, prices increase for tighter arrival-departure windows. The following property places a constraint on the timing of trades.

**Definition 8 (Late allocations).** *Allocations are late if no buyer receives an item before the critical period $t_c(a_i, d_i, b, v_{-i})$, and no seller receives payment before the critical period $t_c(a_i, d_i, s, v_{-i})$.*

Here, the *critical period*, $t_c(a_i, d_i, r_i, v_{-i})$ is the first period, $t' \in [a_i, d_i]$, for which $p_i(a_i, t', r_i, v_{-i}) = \min_{t'' \in [a_i, d_i]}\{p_i(a_i, t'', r_i, v_{-i})\}$, i.e. the first period in which the price is minimal (with respect to $a_i$, $d_i$ and $r_i$). Thus, the *late allocations* property places a timing constraint on the completion of a trade.

**Theorem 2.** *[9] An online auction $M = (g, \tilde{p})$ is truthful if and only if it is implemented by a price-based auction for a **monotonic** and **value-independent** price schedule, $p$, and if it also satisfies **late allocations**.*

Thus, to establish truthfulness we need to establish that the price-ranked DAs are implementing an allocation rule that is admissible with respect to some monotonic and value-independent price schedule, and that the timing satisfies the *late allocations* property.

**Theorem 3.** *A price-ranked DA $M_{\text{pr}}(f)$ is truthful for price schedules that satisfy properties (B1), (B2) and (B3).*

First, note that an agent cannot usefully misrepresent as an agent on the other side of the market. For a buyer, it cannot be a seller because it has no item to offer. A seller on the other hand has no value for acquiring an item. The proof of Theorem 3 then follows from a series of technical lemmas. Taken together, the lemmas imply the existence of a value-independent price schedule $p_i(a_i, d_i, r_i, v_{-i})$ that is monotonic. Notice that we have the *late allocations* property by construction, through the timing of trades (see Section 3).

**Lemma 1.** *If an agent is matched in a price-ranked DA in period $t$, then for fixed $a_i$, $d_i$ and $v_{-i}$ (and random draws in the case of tie-breaking) the price is independent of its reported $w_i$.*

*Proof.* By induction on the period $t$ since an agent's arrival $a_i$, with the inductive hypothesis that either the agent has not yet matched, or if it has matched it has matched at a value-independent price. The base case holds immediately. The inductive case is proved by showing that conditioned on the match occurring in period $t \in [a_i, d_i]$, then it is matched at a value-independent price. For the bid to be matched, it must not be priced-out and so we must have $f_i(t', r_i, v_{-i}) \leq w_i$ for all periods $t' \in [d_i - k, \ldots, t]$, such that $\tilde{ps}_i(t) \leq w_i$. Now, agent $i$ cannot influence the price $f_j$ for any agent $j \neq i$ either now or in any previous period (by B3), and so the provisional prices $\tilde{ps}_j(t)$ to other agents are agent-independent and whether or not other agents are priced-out is also agent-independent (by B3). This implies that the budget-balance check in the price-ranked matching function is independent of agent $i$'s reported value. Finally, the ranking of bids and asks when matching depends on $f_i$ and the final price, $\tilde{ps}_i(t)$, upon matching are both value independent. To handle the case of tie-breaking, just assume that the random draws are agent-independent. □

**Lemma 2.** *If an agent with reported value $w_i$ is not matched in a price-ranked DA, then for fixed $a_i$, $d_i$, $v_{-i}$ (and random draws in the case of tie-breaking), any $w'_i > w_i$ for which would be matched would result in a payment greater than $w_i$.*

*Proof.* Fix $a_i$, $d_i$, $v_{-i}$ and $w_i$, and assume that the bid is not matched. We consider two cases. In each case we argue that any misreport would lead to a price greater than $w_i$ when the agent begins to match.

(Case 1) Bid $i$ never met the budget-balance (BB) condition, but was active with $\tilde{ps}_i(t) \leq w_i$ for all $t \in [a_i, d_i]$. But, agent $i$ cannot influence the BB condition because this depends on its simple-price $f_i(t)$ which is agent-independent and on the prices on the other side of the market. But, while bid $i$ it cannot influence these prices (by B3).

(Case 2) Bid $i$ never met the BB condition while active, but from some period $t$ onwards $\tilde{ps}_i(t) > w_i$. In this case, decreasing $w_i$ has no effect other than to make the bid inactive earlier. Increasing $w_i$ may allow the bid to remain active for more periods and perhaps meet the BB condition and match. But, its payment $\tilde{ps}_i(t')$ in the period $t' \geq t$ in which this occurs satisfies $\tilde{ps}_i(t') > w_i$, since $\tilde{ps}_i(t')$ *increases* with time. □

**Lemma 3.** *Any time agent $i$ with arrival $a_i$ and departure $d_i$ is matched its price is independent of its bid price $w_i$, given properties (B1)–(B3).*

*Proof.* From Lemma 1 we know that if an agent is matched in period $t$, then for fixed $a_i$, $d_i$ and $v_{-i}$ the price is independent of its reported $w_i$. Also, we know that this price is no greater than the minimal $w_i$ for which a match occurs (because otherwise a bid would be priced out). So, the price must be less than or equal to the minimal bid price across all bids that will match. Then, from Lemma 2 we see that this price must be greater than all prices $w_i$ on bids that do not match. Putting this together, we see that there is a value-independent price for trade in period $t$, that is the *minimal bid price that an agent could have bid and traded in that period.* Finally, note that for fixed $[a_i, d_i]$ and fixed $v_{-i}$ the agent will always match in this period for all bid values greater than this price, and so we can associate this price with the price $ps_i(a_i, d_i, r_i, v_{-i})$ as required for the price-based characterization. □

It should also be clear from this discussion that the agent is matched, at this payment, if and only if its bid price is greater than or equal to this price. The remaining condition in the price-based characterization that we must establish is that of monotonicity. We establish this through Lemma 4.

**Lemma 4.** *Fix $w_i, v_{-i}$. The payment $\tilde{ps}_i(t)$ increases with $[a'_i, d'_i] \subseteq [a_i, d_i]$ and an agent cannot match in an earlier period by reporting $[a'_i, d'_i] \subseteq [a_i, d_i]$.*

*Proof.* First, notice that $\tilde{ps}_i(t)$ defined in Equation (3) is independent of $a_i$ and *increases* with an earlier departure. So, the only thing left to show is that an agent cannot bring its match forward by reporting $[a', d'] \subseteq [a, d]$. Notice that it would not be useful to delay its match because $\tilde{ps}_i(t)$ is (weakly) increasing with $t$. Now, announcing an earlier departure does not change the state of the auction in any round $t' \leq d'$ unless the agent is priced-out (because its provisional price becomes higher). So, we must just consider the case of an agent that delays its arrival. But, the only effect is that an agent might forfeit an opportunity to trade. Notice that a later arrival does not change its price schedule of provisional price and so we can restrict attention to possible changes in the matching procedure. But, the agent's report has no effect on its prices or the prices of other agents while it might still trade (B3). So, the only effect could be to cause an ask that would have matched with agent $i$ to match in a period before $i$'s delayed arrival. This only reduces the opportunities to agent $i$ to match once it has arrived. □

## 5 EXPERIMENTAL RESULTS

In this section, we evaluate in simulation each of the price-ranked DAs introduced in Section 3.1. We measure the average value for the schedule (in comparison with the optimal offline solution), as well as the revenue collected by the auctioneer (in comparison with the value of the optimal offline solution). We find that under stochastically stable demand the fixed-price schedule is most efficient, but that this is unable to support efficient matching when demand is more volatile.

We use a commercial integer linear programming algorithm (CPLEX) to compute the offline optimal solution. In determining the offline solution we enforce the constraint that a trade can only be executed if both buyer and seller are present in the same period. The integer program formulation to maximize total value across all feasible schedules is:

$$\max \sum_{(i,j) \in overlap} x_{ij}(w_i + w_j)$$
$$\text{s.t. } 0 \leq \sum_{i:(i,j) \in overlap} x_{ij} \leq 1, \ \forall j \in ask$$
$$0 \leq \sum_{j:(i,j) \in overlap} x_{ij} \leq 1, \ \forall i \in bid$$
$$x_{ij} \in \{0, 1\}, \ \forall i, j,$$

where $(i, j) \in overlap$ is a bid-ask pair that could potentially trade because they have overlapping arrival and departure intervals. Decision variable $x_{ij} \in \{0, 1\}$ indicates that bid $i$ matches with ask $j$.

For the online pricing methods, the simulation performs several training runs to learn the optimal fixed-price and smoothing parameters before calculating market efficiency.

### 5.1 EXPERIMENTAL SET-UP

For each experiment, the simulation runs 50 trials to estimate the mean efficiency. The experiments frequently require additional trials to learn parameters, so there are 16 additional trials for each iteration of the parameter search process. To compare the different price schedules, the simulation uses the same sets of bids and asks to drive each matching method.

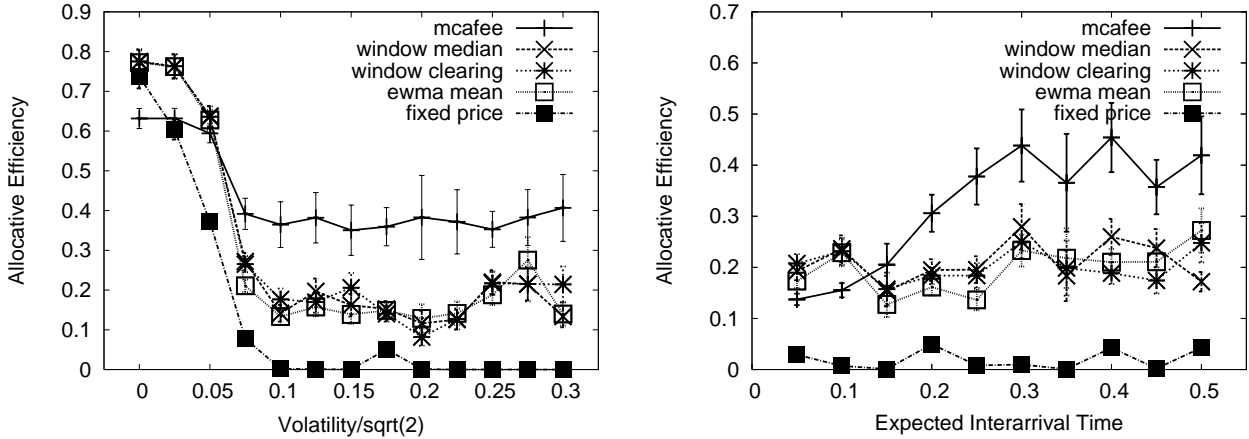

Figure 1: **Mean allocative efficiency plotted against value volatility and arrival intensity. We also plot the 95% confidence intervals.**

Bids and asks are simulated to arrive to the market using a Poisson schedule with fixed intensity until the simulation creates 2000 bids and 2000 asks. Offer durations are sampled from a truncated-exponential distribution. The distribution is truncated at the maximal patience value, with the truncation defined so that 95% of the exponential distribution fits within the patience bound $[0, K]$. To compare trials using different patience bounds, we scale the arrival and departure times to ensure that offers in each experiment have the same expectation for possible trading partners.

The simulation uses a uniform distribution for buyer and seller valuations. In each trial, the simulation initializes the distribution to $U[0, 2]$. After after each time period, the valuation mean either increases or decreases by a relative amount fixed within the trial according to unbiased geometric Brownian motion [5]. The variance of values remains fixed at $\sqrt{2}$ during each time period.

### 5.2 RESULTS

Figure 1 plots the *allocative surplus*[9] relative to the optimal as a function of the inter-period valuation volatility[10] and expected time between offer arrivals. The simulation sets the patience bound for both sets of experiments to $K = 5$. When varying the volatility, the inter-arrival delay between offers is 0.25. The volatility is $\sqrt{2}/10$ per period when inter-arrival time is varied. We observe that when valuations do not change, the fixed-price schedule performs reasonably well. However, as volatility increases or as the market volume decreases (i.e. as the expected inter-arrival time increases) the McAfee method outperforms the other mechanisms. The ewma and windowed schemes perform almost identically and match offers well when volatility is mild or when volume is high. When varying the maximal patience, $K$, we find little, if any, relationship between $K$ and matching efficiency.

---

[9]Defined as the mean ratio of total value from trade with total value from trade computed offline.

[10]Volatility is the standard deviation. We plot against step size on the axis– a constant factor $\sqrt{2}$ less than the actual volatility.

Each of the price schedules, except fixed-price, require that the auctioneer extract revenue from the market. In Table 1, we present the mean and standard deviation of revenue kept by the auctioneer, averaged across all arrival intensities and for low and high volatility trials. While the surplus extracted by the auctioneer can be significant, under most conditions the amount does not change the relative rankings of efficiency if one were to consider only surplus delivered to traders.

## 6 DISCUSSION

We have presented a class of parameterized truthful online DAs. The class is instantiated on fixed-price, moving-average and McAfee-style clearing prices. The biggest outstanding theoretical question is to understand whether our class of DAs includes *all* truthful online DAs. Progress in the characterization of necessary properties would also drive progress in relaxing (or solidifying) some of the current design decisions.

While the general framework is reasonably efficient, further tuning seems possible. For example, matching sorts bids by their simple price-schedules, $f_i$, which underestimates the provisional price, $\tilde{ps}_i$. To improve the situation, we need a better estimator of the provisional price that is agent-independent. We can also explore meta-schemes that track multiple counter-factual worlds with different price schedules and switch between price schedules according to performance while maintaining truthfulness.

|              | Volatility           |          |          |          |
|              | LOW ($\sqrt{2}/40$)  |          | HIGH ($\sqrt{2}/10$) |          |
|              | mean     | stdev    | mean     | stdev    |
|--------------|----------|----------|----------|----------|
| mcafee       | 11.8%    | 1.2%     | 5.1%     | 1.4%     |
| window clear | 3.5%     | 2.0e-2%  | 4.8%     | 6.9e-2%  |
| window median| 3.4%     | 2.1e-2%  | 6.7%     | 1.3e-1%  |
| ewma mean    | 2.9%     | 2.0e-3%  | 7.1%     | 2.8e-2%  |
| fixed price  | 0.0%     | 0.0%     | 0.0%     | 0.0%     |

Table 1: **Expected auctioneer revenue, normalized by the optimal value from trade and averaged across all arrival intensities for low and high value volatility.**

A number of generalizations to our model seem worthy of study. Our current favorites include allowing for heterogeneity in goods and agents, and also allowing for richer models of agent patience.

### Acknowledgments

This work is supported in part by NSF grant IIS-0238147.

### References

[1] Moshe Babaioff, Noam Nisan, and Elan Pavlov. Mechanisms for a spatially distributed market. In *Proc. ACM Conf. on Electronic Commerce*, pages 9–20, 2004.

[2] Moshe Babaioff and William E. Walsh. Incentive-compatible, budget-balanced, yet highly efficient auctions for supply chain formation. In *ACM Conf. on Electronic Commerce*, pages 64–75, 2003.

[3] Avrim Blum, Tuomas Sandholm, and Martin Zinkevich. Online algorithms for market clearing. In *Proc. ACM-SIAM Symp. on Discrete Algorithms*, pages 971–980, 2002.

[4] K Chatterjee and L Samuelson. Bargaining with two-sided incomplete information: An infinite horizon model with alternating offers. *Review of Economic Studies*, 54:175–192, 1987.

[5] Thomas E. Copeland and J. Fred Westion. *Financial Theory and Corporate Policy*. Addison-Wesley, Reading, MA, third edition, 1992.

[6] Rajarshi Das, James E. Hanson, Jeffrey O. Kephart, and Gerald Tesauro. Agent-human interactions in the continuous double auction. In *Proc. International Joint Conf. on Artificial Intelligence*, pages 1169–1187, 2001.

[7] K Deshmukh, Andrew V Goldberg, J D Hartline, and A R Karlin. Truthful and competitive double auctions. In *Proc. European Symp. on Algorithms*, 2002.

[8] Eric Friedman and David C. Parkes. Pricing WiFi at Starbucks– Issues in online mechanism design. In *Proc. ACM Conf. on Electronic Commerce*, pages 240–241, 2003.

[9] Mohammad Taghi Hajiaghayi, Robert Kleinberg, Mohammad Mahdian, and David C. Parkes. Online auctions with re-usable goods. In *Proc. ACM Conf. on Electronic Commerce*, 2005.

[10] Mohammad Taghi Hajiaghayi, Robert Kleinberg, and David C. Parkes. Adaptive limited-supply online auctions. In *Proc. ACM Conf. on Electronic Commerce*, pages 71–80. ACM Press, 2004.

[11] Pu Huang, Alan Scheller-Wolf, and Katia Sycara. Design of a multi-unit double auction e-market. *Computational Intelligence*, 18:596–617, 2002.

[12] R. Lavi and N.Nisan. Online ascending auctions for gradually expiring goods. In *In Proc. ACM-SIAM Symp. on Discrete Algorithms*, 2005.

[13] Ron Lavi and Noam Nisan. Competitive analysis of incentive compatible on-line auctions. In *ACM Conf. on Electronic Commerce*, pages 233–241, 2000.

[14] R. Preston McAfee. A dominant strategy double auction. *Journal of Economic Theory*, 56(2):434–450, April 1992.

[15] Roger B. Myerson and Mark A. Satterthwaite. Efficient mechanisms for bilateral trading. *Journal of Economic Theory*, 29:265–281, 1983.

[16] Ryan Porter. Mechanism design for online real-time scheduling. In *Proc. ACM Conf. on Electronic Commerce (EC'04)*, 2004.

[17] Mark A Satterthwaite and Steven R Williams. Bilateral trade with the sealed bid $k$-double auction: Existence and efficiency. *Journal of Economic Theory*, 48:107–133, 1989.

[18] Vernon L. Smith. An experimental study of competitive market behavior. *Journal of Political Economy*, 70(3):111–137, April 1962.

[19] Gerald Tesauro and Jonathan Bredin. Strategic sequential bidding in auctions using dynamic programming. In *Proc. International Joint Conf. on Autonomous Agents and Multiagent Systems*, pages 591–598, Bologna, Italy, July 2002.

[20] Makoto Yokoo, Y Sakurai, and S Matsubara. The effect of false-name bids in combinatorial auctions: New Fraud in Internet Auctions. *Games and Economic Behavior*, 46(1):174–188, 2004.

[21] Kiho Yoon. The modified vickrey double auction. *Journal of Economic Theory*, 101:572–584, 2001.